\documentclass{article}
\usepackage{times}
\usepackage{amsfonts}
\usepackage[pdfmark,colorlinks]{hyperref}
\begin{document}
\title{Moduli of Quanta}
\author{Jos\'e M. Isidro\\
Instituto de F\'{\i}sica Corpuscular (CSIC--UVEG)\\
Apartado de Correos 22085, Valencia 46071, Spain\\
{\tt jmisidro@ific.uv.es}}

\maketitle

\begin{abstract}
\noindent
The classical phase of the matrix model of 11--dimensional M--theory is complex, infinite--dimensional Hilbert space. As a complex manifold, the latter admits a continuum of nonequivalent, complex--differentiable structures that can be placed in 1--to--1 correspondence with families of coherent states in the Hilbert space of quantum states. The moduli space of nonbiholomorphic complex structures on classical phase space turns out to be an infinite--dimensional symmetric space. We argue that each choice of a complex differentiable structure gives rise to a physically different notion of an elementary quantum.

\end{abstract}

\tableofcontents

\section{Introduction}\label{labastidahijoputa}

String theory defines a quantum--mechanical extension of 10--dimensional supergravity. Now strings have evolved into 11--dimensional M--theory. In turn, M--theory has altered our view not only of strings, from which it derives, but also of quantum mechanics \cite{VAFA}. 
One reason for this is that M--theory itself can be understood as a quantum mechanics. Specifically, the sector of the discrete light--cone quantisation of uncompactified M--theory with momentum $P_-=N/R$ is given by the supersymmetric quantum mechanics of $U(N)$ matrices \cite{BFSS, JAP}. In temporal gauge the action reads
\begin{equation}
S=\frac{1}{2R}\int {\rm d}t\,{\rm tr}\left(\dot X^{\mu}\dot X_{\mu}+\sum_{\mu >\nu}[X^{\mu},X^{\nu}]^2+{\rm i}\theta^t\dot\theta-\theta^t\Gamma_{\mu}[X^{\mu}, \theta]\right),
\label{ramalloketefollen}
\end{equation}
where $\mu,\nu=1,\ldots, 9$ and $R$ is the radius of the 11th dimension of spacetime. Eventually the limits $N\to\infty$, $R\to\infty$ are taken while holding the ratio $P_-=N/R$ fixed (for reviews see, {\it e.g.}, refs. \cite{NICOLAI}, \cite{DEWIT}).

In ref. \cite{MQM} we have studied the truncation of the action (\ref{ramalloketefollen}) obtained by requiring $N$ to be finite and setting the fermions to zero. Then the corresponding classical phase space ${\cal C}$ is given by the product of 9 copies of $\mathbb{C}^{N^2}$, one for each light--cone coordinate in the adjoint representation of $U(N)$. As we let $N\to\infty$, classical phase space becomes infinite--dimensional, separable Hilbert space ${\cal H}_{\cal C}$. The subindex ${\cal C}$ distinguishes this copy of Hilbert space from the Hilbert space of quantum states ${\cal H}_{\cal Q}$ to be introduced presently. In this letter we will analyse the dependence, on the complex moduli of classical phase space ${\cal H}_{\cal C}$, of the quantum mechanics based on the matrix action (\ref{ramalloketefollen}), when one passes to the M--theory limit $N\to\infty$.  Related issues have been analysed in refs. \cite{MATONE, MEX, CARROLL, MARMO}.

\section{Coherent states on Hilbert space}\label{labastidamecagoentuputasombra}

Let us initially set all fermions in eqn. (\ref{ramalloketefollen}) to zero; we will restore them presently.
Picking an orthonormal basis ${\bf e}_j$, $j=1,2,\ldots$ within ${\cal H}_{\cal C}$, any point ${\bf v}$ in classical phase space ${\cal H}_{\cal C}$ can be expressed as
\begin{equation}
{\bf v}=\sum_{j=1}^{\infty}z^j{\bf e}_j, 
\label{ramallohijoputa}
\end{equation}
where the $z^j\in\mathbb{C}$ satisfy
\begin{equation}
\sum_{j=1}^{\infty}\vert z^j\vert^2<\infty.
\label{ramallokasposo}
\end{equation}
Let $\mathbb{C}_{(j)}$ denote the $j$--th complex dimension within ${\cal H}_C$. For every fixed value of $j$, with $j=1,2,\ldots$ the analytic coordinate $z^j$ endows $\mathbb{C}_{(j)}$ with a complex--differentiable structure. The full Hilbert space ${\cal H}_{\cal C}$ can be regarded as 
\begin{equation}
{\cal H}_{\cal C}=\oplus_{j=1}^{\infty}\mathbb{C}_{(j)}
\label{labastidaketefollen}
\end{equation}
subject to eqn. (\ref{ramallokasposo}). The set of all $z=(z^1,z^2, \ldots)$ endows ${\cal H}_{\cal C}$ with an infinite--dimensional complex--differentiable structure, with $z$ as its analytic coordinate (for an introduction to the theory of infinite--dimensional complex manifolds see, {\it e.g.}, ref. \cite{NACHBIN}). The space ${\cal H}_{\cal C}$ carries the Hermitian metric
\begin{equation}
g=\sum_{j=1}^{\infty}{\rm d}\bar z^j{\rm d}z^j.
\label{mekagoentuputakaraluisibanez}
\end{equation}
Let $U({\cal H}_{\cal C})$ denote the unitary group of ${\cal H}_{\cal C}$ with respect to the metric (\ref{mekagoentuputakaraluisibanez}). $U({\cal H}_{\cal C})$ is an infinite--dimensional Lie group \cite{HARPE, KAC} (for a modern, physics--oriented presentation see ref. \cite{THIRRING}).

As a {\it classical}\/ phase space, however, ${\cal H}_{\cal C}$ should be coordinatised by canonical coordinates $q=(q^1,q^2,\ldots)$ and their conjugate momenta $p=(p_1,p_2,\ldots)$. This is achieved by defining $q^j$ and $p_j$ as the real and imaginary parts of $\sqrt{2}z^j$, 
\begin{equation}
z^j=\frac{1}{\sqrt{2}}\left(q^j+{\rm i}p_j\right), \qquad j=1,2,\ldots
\label{labastidaquetepartaunrayo}
\end{equation}
As a real manifold, ${\cal H}_{\cal C}$ has the Darboux coordinates $q^j$, $p_j$, the symplectic form
\begin{equation}
\omega=\sum_{j=1}^{\infty}{\rm d}q^j\wedge{\rm d}p_j,
\label{hijoputaluisibanez}
\end{equation}
and the Euclidean metric 
\begin{equation}
g=\frac{1}{2}\sum_{j=1}^{\infty}\left(({\rm d}q^j)^2+({\rm d}p_j)^2\right).
\label{putoluisibanez}
\end{equation}
Let $O({\cal H}_{\cal C})$ denote the orthogonal group of ${\cal H}_{\cal C}$ with respect to the metric (\ref{putoluisibanez}). $O({\cal H}_{\cal C})$ is also an infinite--dimensional Lie group. 

The $q^j$, $p_j$ above are Darboux coordinates for a harmonic oscillator on $\mathbb{C}_{(j)}$. In fact, setting the fermions to zero and 
expanding the commutator $[X^{\mu}, X^{\nu}]$ in terms of $U(N)$ structure constants, the action (\ref{ramalloketefollen}) reduces to a collection of coupled oscillators. Coupled or independent, the fact is that the corresponding classical phase space is (a number of copies of) the complex plane $\mathbb{C}$. The quantum operators $Q^j$, $P_j$ corresponding to $q^j$, $p_j$ satisfy 
\begin{equation}
[Q^j,P_l]={\rm i}\delta^{j}_{l}\qquad j,l=1,2\ldots
\label{kabronluisibanez}
\end{equation}
These operators act on a Hilbert space of quantum states ${\cal H}_{\cal Q}$. The latter is another copy of complex infinite--dimensional, separable Hilbert space. Although as such it is isomorphic with classical phase space ${\cal H}_{\cal C}$, the subindex ${\cal Q}$ in ${\cal H}_{\cal Q}$ reminds us that we are now dealing with {\it quantum}\/ states. On ${\cal H}_{\cal Q}$ there act annihilation and creation operators
\begin{equation}
A^j=\frac{1}{\sqrt{2}}\left(Q^j+{\rm i}P_j\right),\qquad
(A^j)^+=\frac{1}{\sqrt{2}}\left(Q^j-{\rm i}P_j\right), \qquad j=1,2,\ldots
\label{luisibanezquetepartaunrayo}
\end{equation}
satisfying
\begin{equation}
[A^j,(A^l)^+]=\delta^{jl}\qquad j,l=1,2,\ldots
\label{luisibanezmekagoentuputakaramarikondemierda}
\end{equation}

Let us first consider the finite truncation of the sum (\ref{labastidaketefollen}) given by replacing ${\cal H}_{\cal C}$ with $\mathbb{C}^n$. In the light--cone gauge, and in the presence of $N$ coincident 0--branes, we would have $n=9N^2$. Eventually we will let $n\to\infty$, which is equivalent to the M--theory limit $N\to\infty$. Working for the moment at finite $n$, the Hilbert space of quantum states ${\cal H}_{\cal Q}(\mathbb{C}^n)$ equals the tensor product of $n$ copies of 0--brane Hilbert spaces, $\otimes_{j=1}^n{\cal H}_{\cal Q}(\mathbb{C}_{(j)})$, which again is isomorphic to ${\cal H}_{\cal Q}$. We will refer to ${\cal H}_{\cal Q}(\mathbb{C}_{(j)})$ as a {\it 1--particle Hilbert space}, the idea being that we can trade the Hilbert space of quantum states of a single 0--brane for that of a harmonic oscillator whose phase space is $\mathbb{C}_{(j)}$.
An {\it $n$--particle coherent state}\/ $\vert z^1,\ldots, z^n\rangle$ is defined as the direct product of $n$ copies of 1--particle coherent states,
\begin{equation}
\vert z^1,\ldots, z^n\rangle=\vert z^1\rangle\otimes\ldots\otimes\vert z^n\rangle,
\label{barbonhijoputa}
\end{equation}
on which the annihilation operator $A^l$ acts as 
\begin{equation}
A^l\vert z^1,\ldots, z^n\rangle=z^l\vert z^1,\ldots, z^n\rangle, \qquad l=1,\ldots, n.
\label{barbonputon}
\end{equation}
The resolution of the identity on ${\cal H}_{\cal Q}$ reads 
\begin{equation}
\int_{\mathbb{C}^n}{\rm d}\mu_{(n)}\,\vert z^1,\ldots, z^n\rangle\langle z^1,\ldots, z^n\vert={\bf 1}_{{\cal H}_{\cal Q}}
\label{marikonbarbon}
\end{equation}
with respect to the integration measure on $\mathbb{C}^n$
\begin{equation}
{\rm d}\mu_{(n)}=\left(\frac{1}{\pi}\right)^n{\rm d}^2z^1\wedge\ldots\wedge{\rm d}^2z^n,
\label{mekagoenelputobarbon}
\end{equation}
where
\begin{equation}
{\rm d}^2z^j = \frac{1}{2}\,{\rm d}q^j\wedge{\rm d}p_j, \qquad j=1, \dots, n.
\label{ramallocomprateunchampuanticaspa}
\end{equation}
An important point to bear in mind is that the choice of a complex--differentiable structure on $\mathbb{C}^n$ with analytic coordinates $z^1, \ldots, z^n$ is equivalent to the choice of a set of creation and annihilation operators $A^+_l$, $A^l$ \cite{MPLA}.

Next we pass to the limit $n\to\infty$, where the classical phase space $\mathbb{C}^n$ becomes Hilbert space ${\cal H}_{\cal C}$.  
A mathematically rigorous formalism of creation and annihilation operators for infinite--dimensional phase spaces was described in refs. \cite{SEGAL, JAFFE}, but the heuristic, physics--oriented description that follows is more suitable for our purposes. This alternative description mimicks the definition of field--theory path integrals as infinite products of finite--dimensional integrals. We have a Hilbert space of quantum states ${\cal H}_{\cal Q}$ where coherent states are defined as an infinite tensor product of 1--particle coherent states,
\begin{equation}
\vert z^1,\ldots, z^l, \ldots, \rangle=\vert z^1\rangle\otimes\ldots\otimes\vert z^l\rangle\otimes\ldots
\label{ramalloketehostien}
\end{equation}
The annihilation operator $A^l$ acts as 
\begin{equation}
A^l\vert z^1,\ldots, z^l,\ldots\rangle=z^l\vert z^1,\ldots, z^l, \ldots\rangle, \qquad l=1,2,\ldots
\label{labastidaketehostien}
\end{equation}
In order to write the resolution of the identity on ${\cal H}_{\cal Q}$ we first consider the integral
\begin{equation}
\int_{{\cal H}_{\cal C}}{\rm d}\mu_{(\infty)}\,\vert z^1,\ldots, z^l,\ldots\rangle\langle z^1,\ldots, z^l,\ldots\vert
\label{labastidamarikonazodemierda}
\end{equation}
with respect to the naive integration measure on ${\cal H}_{\cal C}$ 
\begin{equation}
{\rm d}\mu_{(\infty)}=\prod_{l=1}^{\infty}\left(\frac{1}{\pi}\right)^l{\rm d}^2z^1\wedge\ldots\wedge{\rm d}^2z^l\wedge\ldots
\label{ramallokasposocomprateunchampuantikaspa}
\end{equation}
The integral in eqn. (\ref{labastidamarikonazodemierda}) is actually a path integral. As in any path integral that is worthy of its name, we have to factor out a divergent piece, here given by the product $\prod_{l=1}^{\infty}\left(\frac{1}{\pi}\right)^l$ in the measure ${\rm d}\mu_{(\infty)}$. This notwithstanding, factoring out a divergent normalisation may not be enough in order to properly define the measure, and additional regularisations might be required. We will simply proceed ahead under the assumption that an appropriate measure exists, that we will denote by $D^2 z$, in terms of which the resolution of the identity on ${\cal H}_{\cal Q}$ reads
\begin{equation}
\int_{{\cal H}_{\cal C}}D^2z\,\vert z\rangle\langle z\vert = {\bf 1}_{{\cal H}_{\cal Q}}.
\label{ramalloketefollenporhijoputa}
\end{equation}
A Fock--Bargmann space can now be defined as follows. Its elements are analytic functions $f\colon{\cal H}_{\cal C}\rightarrow\mathbb{C}$ in infinite complex variables $z=(z^1, \ldots, z^l, \ldots)$, the latter square--summable as per eqn. (\ref{ramallokasposo}). The functions $f$ are square--integrable on ${\cal H}_{\cal C}$ with respect to the measure $D^2z\,{\rm e}^{-\sum_{j=1}^{\infty}\bar z^j z^j}=D^2z\,{\rm e}^{-\bar z z}$, the scalar product being
\begin{equation}
\langle f\vert g\rangle = \int_{{\cal H}_{\cal C}}D^2z\,{\rm e}^{-\bar zz}\,\overline{f(z)}g(z).
\label{ramalloeresuncasposodemierda}
\end{equation}
Above, $f(z)=\langle \bar z\vert f\rangle$, the $\vert z\rangle=\vert z^1, \ldots,z^l, \ldots\rangle$ being the coherent states (\ref{ramalloketehostien}). 

Some important points must be mentioned.

Without touching on the fine mathematical details of integration theory on infinite--dimensional Hilbert space ${\cal H}_{\cal C}$, which can be found in refs. \cite{SEGAL, JAFFE}, eqns. (\ref{ramalloketehostien})--(\ref{ramalloeresuncasposodemierda}) above can be given an alternative interpretation. For that we resort to the equivalence, metioned after eqn. (\ref{ramallocomprateunchampuanticaspa}), between families of coherent states and complex--differentiable structures. (We will presently give the expression {\it family of coherent states}\/ a precise meaning). Differentiability on infinite--dimensional complex manifolds reduces to (an infinite number of copies of) the Cauchy--Riemann equations. This latter point is elementary, and it stands on a solid mathematical basis  \cite{NACHBIN}. Then we can turn the argument around and {\it declare}\/ that a family of coherent states $\vert z^1, \ldots, z^j,\ldots\rangle\in{\cal H}_{\cal Q}$ is simply a choice of a complex--differentiable structure with analytic coordinates $z^1, \ldots, z^j, \ldots$ on classical phase space ${\cal H}_{\cal C}$.

In turn, a family of coherent states is equivalent, by eqn. (\ref{labastidaketehostien}), to a choice of annihilation operators $A^j$ for all $j=1,2,\ldots$ Once the $A^j$ are known, so are their adjoints $(A^j)^+$. The complete quantum theory on ${\cal H}_{\cal Q}$ corresponding to the classical phase space ${\cal H}_{\cal C}$ can be expressed in terms of annihilation and creation operators. In particular, the notion of an elementary quantum (obtained by the action of a single creation operator on the vacuum) can be reduced to the choice of a complex--differentiable structure on classical phase space ${\cal H}_{\cal C}$.  Any two given complex--differentiable structures on ${\cal H}_{\cal C}$ that are nonbiholomorphic will fail to agree on the notion of an elementary quantum, because the transformation between them involves both $z$ and $\bar z$. That is, this transformation involves both creation and annihilation operators. Since, by definition, an elementary quantum is the result of the action of a single creation operator on the vacuum, such a transformation cannot lead to equivalent notions of an elementary quantum. We conclude that {\it there is a 1--to--1 correspondence between nonbiholomorphic, complex--differentiable structures on ${\cal H}_{\cal C}$ and nonequivalent definitions of an elementary quantum on ${\cal H}_{\cal Q}$}.

We have used harmonic--oscillator coherent states, but our conclusions hold just as well for the matrix--theory dynamics defined by eqn. (\ref{ramalloketefollen}). This is so because the corresponding phase spaces coincide. The states constructed above are coherent both for the harmonic oscillator and for the matrix dynamics of eqn. (\ref{ramalloketefollen}).

\section{Geometry of moduli space}\label{ramallocabron}

Given that there are as many nonequivalent definitions of an elementary quantum on ${\cal H}_{\cal Q}$ as there are nonbiholomorphic, complex--differentiable structures on ${\cal H}_{\cal C}$, the question arises, how many of the latter exist. Now ${\cal H}_{\cal C}$ has a moduli space ${\cal M}({\cal H}_{\cal C})$ of nonbiholomorphic complex structures. It is the symmetric space
\begin{equation}
{\cal M}({\cal H}_{\cal C})=O({\cal H}_{\cal C})/U({\cal H}_{\cal C}).
\label{amico}
\end{equation}
Here the embedding of $U({\cal H}_{\cal C})$ into $O({\cal H}_{\cal C})$ is given by
\begin{equation}
A+{\rm i}B\longrightarrow\left(\begin{array}{cc}
A&B\\
-B&A
\end{array}\right),
\label{labastidaquetelametanporculo}
\end{equation}
where $A+{\rm i}B\in U({\cal H}_{\cal C})$ with $A, B$ real matrices on ${\cal H}_{\cal C}$ \cite{HELGASON}. Let us see how the symmetric space (\ref{amico}) appears as a moduli space of nonequivalent complex structures. Consider the Euclidean metric $g$ of eqn. (\ref{putoluisibanez}). In the complex coordinates of eqn. (\ref{labastidaquetepartaunrayo}), $g$ becomes the Hermitian metric (\ref{mekagoentuputakaraluisibanez}). Now every choice of orthogonal axes $x^j$, $y^j$ in ${\cal H}_{\cal C}$, {\it i.e.}, every element of $O({\cal H}_{\cal C})$, defines a complex structure on ${\cal H}_{\cal C}$ upon setting 
\begin{equation}
w^j=\frac{1}{\sqrt{2}}\left(x^j+{\rm i}y^j\right),\qquad j=1,2,\ldots
\label{kakaxramallo}
\end{equation}
Generically the $w^j$ are related nonbiholomorphically with the $z^j$, because the orthogonal transformation 
\begin{eqnarray}
z^j\longrightarrow w^j&=&\sum_{m=1}^{\infty}\left(R^j_mz^m+S^j_{m}\bar z^{m}\right)\nonumber\\
\bar z^{j}\longrightarrow \bar w^{j}&=&\sum_{m=1}^{\infty}\left(\bar R^{j}_{ m}\bar z^{m}+\bar S^{j}_{m} z^{m}\right),
\label{mierdaxramallo}
\end{eqnarray}
while satisfying the orthogonality conditions
\begin{equation}
\sum_{j=1}^{\infty}\left(R^j_m\,\bar R^{j}_{ n}+S^j_{ n}\, \bar S^{j}_m\right)=\delta_{mn},\qquad
\sum_{j=1}^{\infty}R^j_m\,\bar S^{j}_n=0=\sum_{j=1}^{\infty}S^j_{ m}\,\bar R^{j}_{ n},
\label{kakaxbarbon}
\end{equation}
need not satisfy the Cauchy--Riemann conditions in infinite complex dimensions \cite{NACHBIN}
\begin{equation}
\frac{\partial \bar w^{j}}{\partial z^m}=\bar S^{j}_m=0=S^j_{m}=\frac{\partial  w^{j}}{\partial \bar z^{m}}.
\label{mierdaxcesargomez}
\end{equation}
However, when eqn. (\ref{mierdaxcesargomez}) holds, the transformation (\ref{mierdaxramallo}) is not just orthogonal but also unitary. 
Therefore one must divide the orthogonal group $O({\cal H}_{\cal C})$ by the unitary group $U({\cal H}_{\cal C})$ in order to obtain the parameter space for rotations that truly correspond to inequivalent complex structures on ${\cal H}_{\cal C}$. Nonbiholomorphic complex structures on ${\cal H}_{\cal C}$ are 1--to--1 with rotations that are {\it not}\/ unitary transformations.

The geometry of the moduli space ${\cal M}({\cal H}_{\cal C})$ has been studied in ref. \cite{HARPE}, which the interested reader can consult for details. Here we will just state that ${\cal M}({\cal H}_{\cal C})$ as given by eqn. (\ref{amico}) is a Hilbert manifold and a symmetric space of compact type \cite{HELGASON}. Its connected component has a Poincar\'e polynomial $P(t)$ given by
\begin{equation}
P(t)=\prod_{j=1}^{\infty}\left(1+t^{2j}\right).
\label{ketefollenramallomarikondemierda}
\end{equation}
In terms of Jacobi theta--functions it holds that
\begin{equation}
\prod_{j=1}^{\infty}\left(1+t^{2j}\right)^{24}=\frac{1}{256t^2}\left[\frac{\theta_2(0,t)}{\theta_3(0,t)}\right]^8\left[\frac{\theta_3(0,t)}{\theta_4(0,t)}\right]^4.
\label{ramallokabronhijoputa}
\end{equation}

\section{Discussion}\label{kakaparaluisibanez}

In order to compare our results with those of ref. \cite{PERELOMOV} we first need to truncate the action (\ref{ramalloketefollen}) to a finite $n$, as in ref. \cite{MQM}. Then (bosonic) coherent states $\vert z^1,\ldots, z^n\rangle$ are points on the symmetric space $Sp(2n)/U(n)$ \cite{PERELOMOV}.  Different points on  $Sp(2n)/U(n)$ correspond to coherent states that are connected by means of a nonunitary, symplectic transformation of $\mathbb{R}^{2n}$. On the other hand our conclusions read, for finite $n$,\\
{\it i)} there is a bijection between families of (bosonic) coherent states $\vert z^1,\ldots, z^n\rangle$ and nonbiholomorphic, complex--differentiable structures on classical phase space;\\
{\it ii)} $SO(2n)/U(n)$ is the moduli space for nonbiholomorphic, complex--differentiable structures on $\mathbb{R}^{2n}$.\\
That our results are compatible with those of ref. \cite{PERELOMOV} can be seen as follows. In \cite{PERELOMOV}, (bosonic) coherent states on $\mathbb{R}^{2n}$ are constructed after picking a complex--differentiable structure $J$ on $\mathbb{R}^{2n}$, which is then kept fixed throughout. Associated with this complex structure $J$ there is a whole $Sp(2n)/U(n)$'s worth of coherent states $\vert z^1,\ldots, z^n\rangle$. (This copy of $Sp(2n)/U(n)$ was called a {\it family of coherent states}\/ above).
A different complex structure $\tilde J$ on the same $\mathbb{R}^{2n}$ would give rise to another family of coherent states $\vert \tilde z^1,\ldots, \tilde z^n\rangle$, the latter spanning another copy of $Sp(2n)/U(n)$. The transformation between the $z^j$ and the $\tilde z^l$ on $\mathbb{R}^{2n}$ is nonholomorphic, {\it i.e.}, $\partial \tilde z^k/\partial\bar z^j\neq 0$. This expresses the fact that the families of coherent states respectively represented by $\vert z^1,\ldots, z^n\rangle$ and $\vert \tilde z^1,\ldots, \tilde z^n\rangle$ correspond to different points on the moduli space $SO(2n)/U(n)$, {\it i.e.}, to nonbiholomorphic complex structures $J$, $\tilde J$ on $\mathbb{R}^{2n}$. The picture that emerges is that of a fibration over $SO(2n)/U(n)$, the fibre at each point being $Sp(2n)/U(n)$, and the structure group being $Sp(2n)$. This bundle may, but need not, be trivial. Nontriviality of this bundle implies that the passage from the quantum theory based on the complex structure $J\in SO(2n)/U(n)$ to the quantum theory based on the complex structure $\tilde J\in SO(2n)/U(n)$ is generally not the identity transformation. Hence the corresponding quantum mechanics do not necessarily agree: this is the notion of a duality transformation. Finally passing to the M--theory limit $n\to\infty$, we have a fibration of $Sp({\cal H}_{\cal C})/U({\cal H}_{\cal C})$ over the base manifold $O({\cal H}_{\cal C})/U({\cal H}_{\cal C})$, with structure group $Sp({\cal H}_{\cal C})$. 

Next turning on the fermions that were so far set to zero, let us call $q_{\theta}=\theta$ in eqn. (\ref{ramalloketefollen}), so we have a conjugate momentum $p_{\theta}={\rm i}\theta^t$. The $\theta$ have a coupling to the $X^{\mu}$ that resembles the usual  minimal coupling to a gauge field. It differs from being minimal coupling by the same structure constants that prevent the potential for the bosonic coordinates $X^{\mu}$ from being the usual harmonic potential. The fermionic piece of phase space is spanned by $q_{\theta}$ and $p_{\theta}$, with a form $\omega_{\theta}={\rm d}q_{\theta}\wedge{\rm d}p_{\theta}$ that is symmetric instead of symplectic because the $\theta$'s are Grassmann. In the quantum theory this leads to anticommutators instead of commutators. Again let us first consider the truncation of (\ref{ramalloketefollen}) to a finite $n$, so that fermionic coherent states can now be constructed as in ref. \cite{PERELOMOV}: given a fixed complex--differentiable structure on $\mathbb{R}^{2n}$, fermionic coherent states span the symmetric space $SO(2n)/U(n)$. Letting now the complex structure vary we obtain a fibration of $SO(2n)/U(n)$ over $SO(2n)/U(n)$, the fibre being a family of fermionic coherent states for a fixed complex structure, the base being the moduli space of complex structures on $\mathbb{R}^{2n}$, and the structure group being $SO(2n)$. In the M--theory limit $n\to\infty$ we obtain a fibration of $O({\cal H}_{\cal C})/U({\cal H}_{\cal C})$ over $O({\cal H}_{\cal C})/U({\cal H}_{\cal C})$ with structure group $O({\cal H}_{\cal C})$.

To summarise, complex--differentiable structures on classical phase spaces ${\cal C}$ have a twofold meaning. Geometrically they define complex differentiability, or analyticity, of functions on complex manifolds such as ${\cal C}$. Quantum--mechanically they define the notion of a quantum, {\it i.e.}, an elementary excitation of the vacuum state. In this letter we have elaborated on this latter meaning. The mathematical possibility of having two or more nonbiholomorphic complex--differentiable structures on a given classical phase space leads to the physical notion of a quantum--mechanical duality, {\it i.e.}, to the relativity of the notion of an elementary quantum. This relativity is understood as the dependence of a quantum on the choice of a complex--differentiable structure on ${\cal C}$. One can summarise this fact in the statement that a quantum is a complex--differentiable structure on classical phase space \cite{GOLM}. A duality arises as the possibility of having two or more, apparently different, descriptions of the same physics. These facts imply that the concept of a quantum is not absolute, but relative to the theory that one measures with \cite{VAFA}. 

\section{Conclusions}

In this letter we have analysed the dependence of the notion of a quantum on the complex--differentiable structure chosen on classical phase space. When the latter is complex infinite--dimensional Hilbert space ${\cal H}_{\cal C}$ have established the existence a moduli space of nonbiholomorphic complex structures ${\cal M}({\cal H}_{\cal C})$ given by the quotient of the orthogonal group by the unitary group, $O({\cal H}_{\cal C})/U({\cal H}_{\cal C})$. Moving around within ${\cal M}({\cal H}_{\cal C})$ we obtain nonequivalent definitions of complex differentiability. The transformations between observers carrying nonbiholomorphic complex structures are nonholomorphic. Hence the corresponding observers do not agree on the notion of an elementary quantum. On top of each point of ${\cal M}({\cal H}_{\cal C})$ there stands a whole family of coherent states given by $Sp({\cal H}_{\cal C})/U({\cal H}_{\cal C})$ for bosonic states and by $O({\cal H}_{\cal C})/U({\cal H}_{\cal C})$ for fermionic states.

The possibility of transforming between nonequivalent definitions of a {\it quantum}, or even between {\it quantum}\/ and {\it classical}, leaves one wondering if, as already stated in ref. \cite{VAFA}, section 6, classicality {\it vs.} quantumness won't be a matter of convention, after all. We have taken our starting point in the matrix theory action (\ref{ramalloketefollen}); it has been argued that matrix quantum mechanics describes the fundamental dynamics underlying M--theory. However our conclusions are not limited to an M--theory context. Thus, {\it e.g.}, whatever one's favourite choice is for a quantum theory of gravity (for recent reviews see, {\it e.g.}, ref. \cite{ASHTEKAR, NPZ}), quantising gravity is dual to relativising the notion of a quantum. In rendering the concept of a quantum observer--dependent we are, in a sense, quantising gravity. Thus, rather than searching for the quanta of gravity, we are searching for the moduli of quanta.

\vskip1cm
{\bf Acknowledgements}

It is a great pleasure to thank J. de Azc\'arraga for encouragement and support. The author thanks Max-Planck-Institut f\"ur Gravitationsphysik (Potsdam, Germany) where this work was begun, for hospitality. This work has been partially supported by EU network MRTN--CT--2004--005104,
by research grant BFM2002--03681 from Ministerio de Ciencia y Tecnolog\'{\i}a, by research grant GV2004--B--226 from Generalitat Valenciana, by EU FEDER funds, by Fundaci\'on Marina Bueno and by Deutsche Forschungsgemeinschaft.

\end{document}